\documentclass[useAMS,usenatbib]{mn2e}

\usepackage{latexsym}
\usepackage{amssymb}
\usepackage{amsfonts}
\usepackage{amsmath}
\usepackage[dvips]{graphicx}
\usepackage{bm}

\title[Gravitational Faraday Rotation in Binary Pulsar Systems]{Gravitational Faraday Rotation in Binary Pulsar Systems}
\author[M.L. Ruggiero and A. Tartaglia]{M. L. Ruggiero$^{1,2}$\thanks{E-mail: matteo.ruggiero@polito.it} and
 A. Tartaglia$^{1,2}$\thanks{E-mail: angelo.tartaglia@polito.it}\\
$^{1}$Dipartimento di Fisica, Politecnico di Torino, Corso Duca degli Abruzzi 23, Torino, Italy\\
$^{2}$INFN, Sezione di Torino, Via Pietro Giuria 1, Torino, Italy}
\begin{document}

%\date{Accepted 1988 December 15. Received 1988 December 14; in original form 1988 October 11}

%\pagerange{\pageref{firstpage}--\pageref{lastpage}} \pubyear{2002}

\maketitle

\label{firstpage}

\begin{abstract}
We study the gravitational Faraday rotation, on linearly polarized light
rays emitted by a pulsar, orbiting another compact object. We relate the
rotation angle to the orbital phase of the emitting pulsar, as well as to
other parameters describing its orbit and the orientation of the angular
momentum of the binary companion. We give numerical estimates of the effect
for the double-pulsar system PSR J0737-3039, and we note that the expected
magnitude is exceedingly small, making the effect unlikely to be observed
with present technology. It is however interesting per se, since in this
phenomenon, gravito-magnetism plays a leading role, unlike what happens, for
instance, when studying light bending or gravitational time delay, where it
appears as a correction to the gravito-electric contribution. Also, we
envisage the possibility that this effect could be relevant, at least in
principle, for a pulsar orbiting a non charged black-hole.
\end{abstract}

%\date{Accepted 1988 December 15. Received 1988 December 14; in original form 1988 October 11}

%\pagerange{\pageref{firstpage}--\pageref{lastpage}} \pubyear{2002}


%----------------------------Section-------------------------------

\section{Introduction}

\label{sec:intro}
%----------------------------Section-------------------------------

The physical properties of the recently discovered double-pulsar system PSR J0737-3039 (\cite{burgay03},
\cite{lyne04}) make this system a rare laboratory for testing relativistic gravity. Namely, the system is
composed of a 22-ms pulsar, PSR J0737-3039 A, in a slightly eccentric ($e \simeq 0.088 $) $2.4 h$ orbit,
together with PSR J0737-3039 B, whose period is 2.7 s: thanks to the high mean orbital speed ($v \simeq 0.001
c$) this system is highly relativistic, and allows the detection of four post-Keplerian parameters, thus
accurately constraining the masses of the two stars and providing also stringent tests of General Relativity
(GR) (\cite{kramer05}).

Actually, excluding cosmological issues, GR has passed all observational
tests with excellent results (\cite{will05}). However, we must remember that
most of the tests of GR come from Solar System experiments, where the
gravitational field is in the "weak" regime. On the other hand, it is
expected that deviations from GR can occur in the "strong" field regime:
hence, the Solar System experiments are inadequate to this end. On the
contrary, the strong gravitational field is best tested by means of pulsars.
Among the post-Newtonian effects that should be relevant in strong field
regime (\cite{will}), there are the so called \textit{gravito-magnetic} (GM)
effects, which arise from the rotation of the sources of the gravitational
field: these effects are generally small, since the gravitational coupling
with the angular momentum of the source is much weaker than the coupling
with mass alone (the so called \textit{gravito-electric} interaction), and
this makes their detection very difficult (see \cite{mashh1,ruggiero02}).

In previous papers (\cite{tartaglia05,ruggiero05}), we pointed out that
binary pulsar systems in general and, in particular, the double-pulsar PSR
J0737-3039, could be very interesting for testing the GM effects. The
relevance of GM effects in binary pulsar systems was investigated in the
past and also more recently by many authors, who also pointed out the
difficulties that the detection of these effects must face (\cite%
{doro95,laguna97,wex99,kopeikin02,rafikov06a,rafikov06b}), mainly due to
their smallness, which is overwhelmed by other dominant effects.

Here, we would like to study a physical phenomenon where the
gravito-magnetic interaction has a leading role, and which can be tested, at
least in principle, in binary pulsar systems, i.e. systems where a pulsar is
orbiting a compact object, such as another pulsar, as in PSR J0737-3039, or,
hopefully, a black hole. The phenomenon we are going to deal with is the
gravitational Faraday rotation.

The polarization vector of an electromagnetic (e.m.) wave
propagating in a gravitational field undergoes a change of
direction, as a result of the deflection of e.m. signals, and, in
addition a rotation around the propagation vector. As for the
latter, in particular, it has been studied and discussed by many
authors, in the context of GR (see, for early works the paper by
\cite{plebanski60} and, for more recent references the paper by
\cite{sereno05a}). Actually this gravitational rotation of the
plane of polarization, for linearly polarized light rays, is the
gravitational analogue of the electro-magnetic Faraday effect,
that is the rotation which a light ray undergoes when it
propagates in a plasma under the action of a magnetic field, the
role of the magnetic field being played, in this case, by the
gravito-magnetic field. It is possible to show that, for
astrophysical interesting cases, a spherical and non rotating
source of gravitational field does not cause a net rotation of the
polarization vector (\cite{plebanski60,dyer92,faraoni93}), while
things change when rotating sources come into play. Indeed, the
gravitational Faraday rotation is a purely gravito-magnetic
phenomenon (\cite{sereno05a}). Actually, also the case of rotation
is somewhat subtle, since in the lowest order of the weak field
approximation, \cite{plebanski60} concluded that the rotation of
the polarization plane occurs only when the light rays penetrate
into a rotating distribution of matter. More accurate
calculations, that take into account the deflection of the light
rays and higher-order terms in the weak field approximation, show
that rotation occurs outside the horizon of a rotating (Kerr)
black hole (\cite{ishi88,nouri99,sereno05a}).

The phenomena we try and study here could be particularly relevant in the case of X-ray emission from accretion
disks or other thick material envelopes surrounding rotating magnetized compact objects and specifically black
holes. Indeed the intrinsic birefringence of the plasma in the magnetic field, together with the general
relativistic effects due to the curvature in proximity of horizons, can be  the primary source of  polarization
and of the rotation of the polarization plane of the emitted short wavelength radiation
(\cite{connors,broderick}).

Under the hypothesis that, to a reasonable order of approximation, the
space-time around a rotating compact object (i.e. a pulsar, or a black hole)
can be described by the Kerr metric, here we investigate the gravitational
Faraday rotation induced on the light rays emitted by a pulsar orbiting
around it. We study the effect and give estimates of its magnitude, and
evaluate the possibility of detecting it.

The paper is organized as follows: after briefly reviewing the basics of propagation of e.m. signals in a
gravitational field that lead to the gravitational Faraday rotation in Sec. \ref{sec:em}, we apply the formalism
to the case of the signals emitted by a pulsar orbiting a compact object in Sec. \ref{sec:gravfar}. The estimate
of the magnitude of the effect, together with the evaluation of the possibility of detecting it, is discussed in
Sec. \ref{sec:disc}, while the conclusions are summarized in Sec. \ref{sec:conc}

%----------------------------Section-------------------------------

\section{Propagation of e.m. signals in a gravitational field and the
gravitational Faraday effect} \label{sec:em}
%----------------------------Section-------------------------------

The propagation of e.m. signals in a gravitational field, can be studied by
writing the Maxwell equations in the Lorentz gauge; on following \cite%
{fayos82}, we may write \footnote{%
Greek indices run from 0 to 3, Latin indices run from 1 to 3; the space-time
metric has signature $(1,-1,-1,-1)$, and we use units such that G=c=1;
boldface letters refer to three dimensional vectors.}:

\begin{equation}
\nabla_\alpha \nabla^\alpha A^\beta = 0,  \label{eq:maxwell11}
\end{equation}
\begin{equation}
\nabla_\alpha A^\alpha = 0.  \label{eq:maxwell12}
\end{equation}

This is done considering two basic approximations: (i) the electromagnetic field is weak enough not to affect
the gravitational field; (ii) the
geometric optics limit is adopted, where it is assumed that the wavelength $%
\lambda$ of the e.m. waves is much smaller than $L$, where $L=min(\mathcal{L}%
,R)$, $\mathcal{L}$ being a typical distance where amplitude, polarization
and wavelength vary appreciably and $R$ is the order of magnitude of the
(spatial) curvature radius. Then, we may introduce the wave four-vector $%
k^\alpha$ and polarization four-vector $f^\alpha$, which fulfill the
following relations
\begin{equation}
k^\alpha k_\alpha=0, \ \ \ k^\alpha f_\alpha=0, \ \ \ f^\alpha f_\alpha=1.
\label{eq:fourvectors1}
\end{equation}

Light rays follow null geodesics, whose tangent vector is $k^\alpha$;
furthermore, the polarization four vector undergoes parallel transport along
these geodesics. Consequently, we may write
\begin{equation}
k^\alpha \nabla_\alpha k^\beta = 0,  \label{eq:geok1}
\end{equation}
\begin{equation}
k^\alpha \nabla_\alpha f^\beta = 0.  \label{eq:geof1}
\end{equation}

The changes induced by the gravitational field on the wave and polarization
four-vectors, are then obtained integrating the equations (\ref{eq:geok1}-%
\ref{eq:geof1}). Actually, this problem can be suitably studied in the 1+3
projection formalism, applied to the stationary space-time describing the
gravitational field in which waves propagate. This approach is outlined by
\cite{fayos82,nouri99}: here, we briefly recall its foundations.

Let $\mathcal{M}$ be the manifold which is the model of our space-time,
endowed with the metric $g_{\alpha\beta}$, together with a metric connection
$\Gamma$. Since the space-time is stationary, there exists a time-like
Killing vector field $\xi_t$. Then, one can show that, for each $p \in
\mathcal{M}$, there is a three-dimensional manifold $\Sigma_3$, defined by
the smooth map

\begin{equation}
\Psi : \mathcal{M} \rightarrow \Sigma_3,  \label{eq:mapmsigma31}
\end{equation}
$\Psi=\Psi(p)$ being the orbit of the Killing field $\xi_t$ passing through $%
p$. We can introduce a coordinate system adapted to the congruence $%
\xi_t=\partial_t$, and let $\gamma_{ij}$ be the projected three-dimensional
metric of $\Sigma_3$. Suitable differential operators can be also defined on
$\Sigma_3$, starting from $\gamma_{ij}$, in such a way that a covariant
derivative $\bm{\nabla}$ for the three vectors belonging to $\Sigma_3$ is
defined (see, for details \cite{fayos82,nouri99,landau} and references
therein).

If we write the metric of a stationary space-time in the form
\begin{equation}
ds^2=h \left(dx^0-A_i dx^i \right)^2-dl^2,  \label{eq:statmet1}
\end{equation}
where
\begin{equation}
A_i \doteq -\frac{g_{0i}}{g_{00}}, \ \ \ h \doteq g_{00},
\label{eq:statmet2}
\end{equation}
and
\begin{equation}
dl^2 = \left(-g_{ij}+\frac{g_{0i}g_{0j}}{g_{00}} \right)dx^i dx^j \doteq
\gamma_{ij} dx^i dx^j,  \label{eq:statmet3}
\end{equation}
it is possible to introduce the gravito-electric and gravito-magnetic
fields, given respectively by
\begin{equation}
\bm{E}_g = -\bm{\nabla} h^{1/2},  \label{eq:defge1}
\end{equation}
\begin{equation}
\bm{B}_g = \bm{\nabla} \times \bm{A}.  \label{eq:defgm1}
\end{equation}

In this context, on the basis of an orthogonal decomposition in adapted coordinates, the three-vectors $\bm{k}$
and $\bm{f}$, representing the projection of $k^\alpha, f^\alpha$ on $\Sigma_3$, can be taken to be equivalent
to the contravariant components of $k^\alpha$ and $f^\alpha$, respectively. Then, from equations
(\ref{eq:geok1}-\ref{eq:geof1}), it is possible to write the three-dimensional equations

\begin{equation}
\bm{\nabla}_{\bm{k}} \bm{k} = \bm{L} \times \bm{k}+\left(\bm{E}_g \cdot %
\bm{k} \right) \bm{k}  \label{eq:geok2}
\end{equation}
\begin{equation}
\bm{\nabla}_{\bm{k}} \bm{f} = \bm{L} \times \bm{f}  \label{eq:geof2}
\end{equation}
where $\bm{L}=\bm{L}\left(\bm{E}_g,\bm{B}_g,\bm{k},\bm{f} \right)$ (see \cite%
{fayos82,nouri99}.

Eqs. (\ref{eq:geok2}-\ref{eq:geof2}) show, in particular, that the
polarization vector rotates with angular velocity $\bm{L}$ along the
projected geodesic. By carrying out explicit calculations, the angle of
rotation around $\bm{k}$ which expresses the gravitational Faraday rotation
is given by
\begin{equation}
\alpha_g = -\frac{1}{2}\int^{obs}_{sou} \sqrt{h} \bm{B}_g \cdot d\bm{l},
\label{eq:omega1}
\end{equation}
where the integral is evaluated from the source (emission point) to the
observer, along the null geodesics whose tangent (spatial) vector is $\hat{%
\bm{k}}$, such that $d\bm{l}=\hat{\bm{k}}dl$.

It is useful to recall here the expression of the electromagnetic Faraday rotation (\cite{kramer}):

\begin{equation}
\alpha_{Faraday} =\frac{4 \pi e^{3}}{m^{2}c^{2}\omega ^{2}}\int_{sou}^{obs}nB_{%
\parallel }dl,  \label{eq:omegaem1}
\end{equation}%
where we have the physical quantities that describe the properties of the
plasma where the wave propagates ($n$ is the refraction index), the angular
frequency $\omega $ of the radiation and $B_{\parallel }$, which is the
component of the magnetic field along the line of sight. On comparing (\ref%
{eq:omega1}) and (\ref{eq:omegaem1}) we see that both depend on the component of the gravito-magnetic or
magnetic field along the line of sight. There is a formal analogy, even though the effects are physically
different since the gravitational Faraday rotation is a purely geometric effect, while the traditional Faraday
rotation depends on the frequency of the light rays.

Formula (\ref{eq:omega1}) shows also that the gravitational Faraday rotation
is an effect due to mass-current, i.e. to the rotation of the source of the
gravitational field, so it is a purely gravito-magnetic phenomenon, and it
is null in space-times around non rotating sources.

As we said before, this effect was investigated in the past by many authors, in suitable weak field
approximations. We are interested in the effect on light rays emitted by a pulsar orbiting a compact object:
hence, we cannot apply here calculations outlined in \cite{plebanski60}, since their level of approximation is
too low, and gives no rotation of the polarization vector unless light rays penetrate into rotating
distributions of matter.

A better level of approximation was considered by \cite%
{ishi88,nouri99,sereno05a}, who studied propagation of light rays outside the horizon of a rotating Kerr black
hole. These authors agree on the right order of approximation needed, however there is some disagreement on
numerical values. In particular, \cite{sereno05a} obtained for the gravitational rotation angle the expression
\begin{equation}
\alpha_g = -\mu \frac{\pi}{4} \frac{M J_\parallel}{\xi^3}  \label{eq:omega2}
\end{equation}
where $\mu$ is a parameter that quantifies the contribution of angular
momentum to space-time curvature ($\mu=1$ in GR), $M$ is the mass of the
source of gravitational field and $J_\parallel$ is the component of its
angular momentum along the line of sigh, $\xi$ is the projection in the
plane of sight of the distance between the emission point and the source of
the gravitational field.

In the following Section, under the hypothesis that, to a reasonable order
of approximation, the space-time around a rotating compact object (i.e. a
pulsar, or a black hole) can be described by the Kerr metric, we apply (\ref%
{eq:omega2}) in order to study the gravitational Faraday rotation on the
light rays emitted by a pulsar orbiting the compact object.

Before going on, we notice that a better (or more realistic) approximation for describing the space-time around
a rotating compact object such as a pulsar, is given by the metric obtained by \cite{hartle68}. Since a neutron
star's moment of inertia is larger than the one of a black hole of similar mass, the Hartle-Thorne metric could
lead to a slightly larger gravitational Faraday rotation, however we do not expect that this can significantly
change the order of magnitude of the effect.

%----------------------------Section-------------------------------

\section{Gravitational Faraday Effect in a binary pulsar system}

\label{sec:gravfar}
%----------------------------Section------------------------------

%----------------------------Figure-------------------------------
\begin{figure}
\begin{center}
\includegraphics[width=9cm,height=9cm]{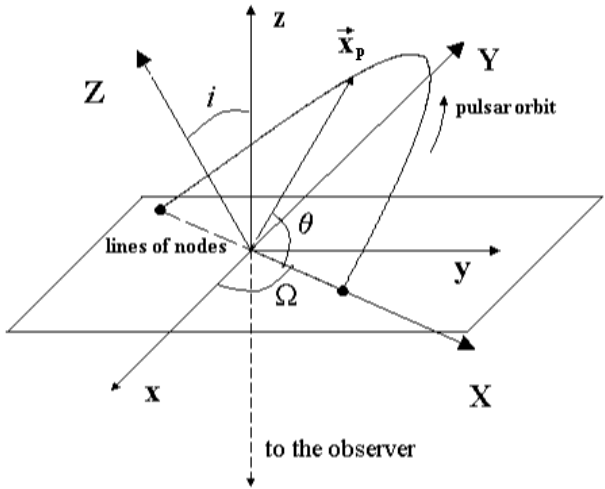}
\end{center}
\caption{{\protect\small Notation used for describing the pulsar orbit.}}
\label{fig:orbita1}
\end{figure}
%----------------------------Figure-------------------------------

We use the notation of Figure \ref{fig:orbita1} for the description of the
pulsar orbit around the compact binary companion (see \cite{straumann}). We
choose a first set of Cartesian coordinates $\{x,y,z\}$, with origin in the
center of mass of the binary system, and such that the line of sight is
parallel to the $z$ axis. Then, we introduce another set of Cartesian
coordinates $\{X,Y,Z\}$, with the same origin: the $X$ axis is directed
along the ascending node, the $Z$ axis is perpendicular to the orbital
plane. The angle between the $x$ and $X$ axes is $\Omega$, the longitude of
the ascending node, while the angle between the $z$ and $Z$ axes is $i$, the
inclination of the orbital plane.

Let $\vec{\bm{x}}_p=r_p \hat{\bm{x}}_p$ be the vector describing the orbit
of the pulsar (here and henceforth, the suffix $p$ refers to the emitting
pulsar, while $c$ refers to its binary companion): it is described by
\begin{equation}
\hat{\bm{x}}_p = \cos(\omega+\varphi) \hat{\bm{X}}+\sin(\omega+\varphi) \hat{%
\bm{Y}},  \label{eq:orbita1}
\end{equation}
in terms of the argument of the periastron, $\omega$, and the true anomaly, $%
\varphi$. Let us pose
\begin{equation}
\theta \doteq \omega+\varphi,  \label{eq:deftheta}
\end{equation}
for the sake of simplicity. Then, we use the notation
\begin{equation}
\vec{\bm{r}} \doteq \vec{\bm{x}}_p-\vec{\bm{x}}_c  \label{eq:defr1}
\end{equation}
to describe the position of the pulsar with respect to its companion, and we
remember that we have, for the Keplerian problem
\begin{equation}
r = \frac{a(1-e^2)}{1+e\cos\varphi},  \label{eq:kep1}
\end{equation}
where $a$ is the semi-major axis of the relative motion and $e$ is the
eccentricity. The astronomical elements $\Omega$, $i$, $\omega$, $a$, $e$
represent the Keplerian parameters.

%----------------------------Figure-------------------------------
\begin{figure}
\begin{center}
\includegraphics[width=6cm,height=6.3cm]{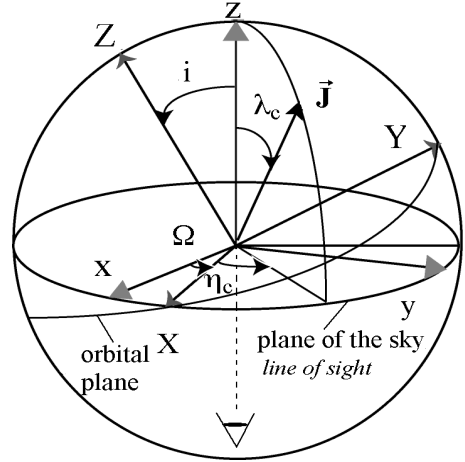}
\end{center}
\caption{{\protect\small Notation and conventions for describing the pulsar
orbit and the spin axis of the binary companion. The spin axis $\vec{\bm{J}}$
orientation is determined by the angles $\protect\lambda_c,\protect\eta_c$.
The system $\{x,y,x\}$ is such that $xy$ is the plane of sight, the $%
\{X,Y,Z\}$ system is such that $XY$ is the orbital plane.}}
\label{fig:orbita_spin}
\end{figure}
%----------------------------Figure-------------------------------

Then, let $\vec{\bm{J}}$ be the angular momentum of the binary companion: its direction is determined by the
angle $\lambda_c \doteq \hat{\bm{z}} \cdot \hat{\bm{J}}$, and $\eta_c$ between the projection of the
$\vec{\bm{J}} $ onto the plane of the sky and the ascending node of the binary orbit (see Figure
\ref{fig:orbita_spin}).

As a consequence, we may write

\begin{equation}
\hat{\bm{J}}=\sin \lambda _{c}\cos \left( \Omega +\eta _{c}\right) \hat{%
\bm{x}}+\sin \lambda _{c}\sin \left( \Omega +\eta _{c}\right) \hat{\bm{y}}%
+\cos \lambda _{c}\hat{\bm{z}}.  \label{eq:proJXYZ1}
\end{equation}

The unit vector along the line of sight is $\hat{\bm{n}}=-\hat{\bm{z}}$, so
that we have
\begin{equation}
J_{\parallel }=|\vec{\bm{J}}|\left( -\cos \lambda _{c}\right) .
\label{eq_defJpar1}
\end{equation}

Furthermore since $\xi=|\hat{\bm{n}}\times \left(\vec{\bm{r}} \times \hat{%
\bm{n}} \right)|$, from (\ref{eq:orbita1}-\ref{eq:defr1}) and
\begin{equation}
\hat{\bm{n}} = -\cos i \hat{\bm{Z}} - \sin i \hat{\bm{Y }} ,
\label{eq:defn1}
\end{equation}
we have
\begin{equation}
\xi= r \sqrt{\cos^2 \theta+\sin^2 \theta \cos^2 i}.  \label{eq:defxi1}
\end{equation}

Consequently, the rotation angle of the polarization vector (\ref{eq:omega2}%
), setting $\mu=1$ as in GR, becomes

\begin{equation}
\alpha _{g}= \frac{\pi }{4}\frac{M|\vec{\bm{J}}|}{r^{3}}\frac{\cos \lambda _{c}}{\left( \cos ^{2}\theta +\sin
^{2}\theta \cos ^{2}i\right) ^{3/2}},  \label{eq:omega3}
\end{equation}%
or, explicitly writing the equation of the orbit (\ref{eq:kep1}) and
re-introducing the mean anomaly

\begin{equation}
\alpha _{g}= \frac{\pi }{4}\frac{G^{2}}{c^{5}}\frac{M|\vec{\bm{J}}|}{%
a^{3}(1-e^{2})^{3}}\frac{\cos \lambda _{c}\left( 1+e\cos \varphi \right) ^{3}%
}{\left( \cos ^{2}\left( \omega +\varphi \right) +\sin ^{2}\left( \omega
+\varphi \right) \cos ^{2}i\right) ^{3/2}},  \label{eq:omega4}
\end{equation}%
in physical units.

We see that the rotation angle depends on the orbital phase: this fact,
could lead, in principle, to a way of detecting it.

%----------------------------Section-------------------------------

\section{Discussion}

\label{sec:disc}
%----------------------------Section-------------------------------

%----------------------------Figure-------------------------------
\begin{figure}
\begin{center}
\includegraphics[width=6cm,height=6cm]{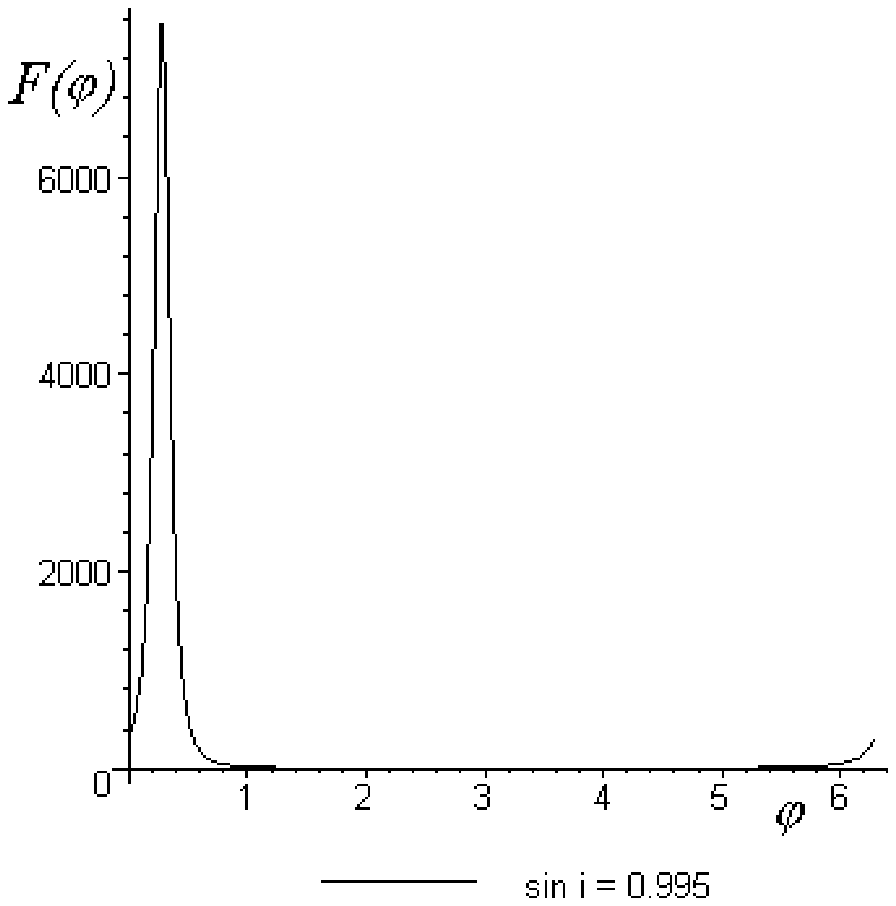} %
\includegraphics[width=6cm,height=6cm]{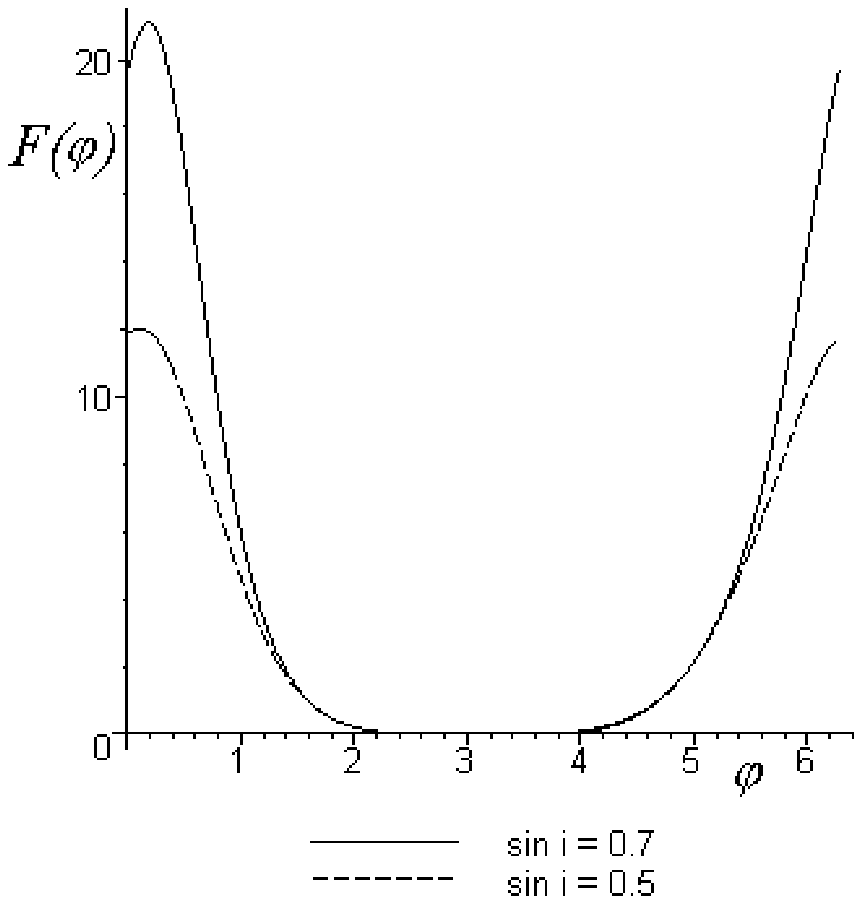}
\end{center}
\caption{{\protect\small Variation of the phase $F(\protect\varphi)$, along the pulsar orbit: above, the
inclination of the orbit is the one measured in the double-pulsar PSR J0737-3039; below, two smaller values have
been chosen.}} \label{fig:fphi}
\end{figure}
%----------------------------Figure-------------------------------

In order to quantitatively evaluate the magnitude of the gravitational
Faraday rotation for light rays emitted by a pulsar in a binary system, it
is useful to write (\ref{eq:omega4}) in the form

\begin{equation}
\alpha _{g}=\mathcal{A}F(\varphi ),  \label{eq:omega5}
\end{equation}%
where
\begin{equation}
\mathcal{A}\doteq  \frac{\pi }{4}\frac{G^{2}}{c^{5}}\frac{M|\vec{\bm{J}}%
|\cos \lambda _{c}}{a^{3}(1-e^{2})^{3}},  \label{eq:defA1}
\end{equation}%
might be thought of as an "amplitude", independent of the orbital phase,
while
\begin{equation}
F(\varphi )\doteq \frac{\left( 1+e\cos \varphi \right) ^{3}}{\left( \cos
^{2}\left( \omega +\varphi \right) +\sin ^{2}\left( \omega +\varphi \right)
\cos ^{2}i\right) ^{3/2}}  \label{eq:defF1}
\end{equation}%
is a "phase" factor, explicitly depending on the anomaly $\varphi $.

For estimating $\mathcal{A}$, we consider, for instance, data coming from
PSR J0737-3039 (\cite{kramer05}): accordingly, $a=9\times 10^{5}\
Km,e=0.088,\omega =75.85^{\circ },\sin i=0.995,M=1.250M_{\odot }$ (we
considers light rays emitted by PSR J0737-3039 A, while it orbits PSR
J0737-3039 B), and estimate $|\vec{\bm{J}}|\simeq |\vec{\bm{J}}_{\odot }|$.
Since we don't know the orientation of the spin of PSR J0737-3039 B, we may
suppose, for the sake of simplicity $\cos \lambda _{c}=1$, which is the most
favorable situation. Consequently, we obtain
\begin{equation}
\mathcal{A}\simeq 0.94\times 10^{-18}  rad. \label{eq:estA1}
\end{equation}

The function $F(\varphi)$ is plotted in Figure \ref{fig:fphi}, for different values of the inclination. We see
that, for orbits with great inclination, the effect grows as far as $\omega+\varphi \rightarrow \pi/2,3\pi/2$,
i.e. close to conjunction.

We may say that, provided that $J_\parallel \neq 0$, the effect is bigger for small orbits with great
eccentricity and inclination. Actually, even though the peculiar dependence on the orbital phase may give a
chance to measure the rotation angle, in very favorable conditions, the overall effect remains very small, and
furthermore, it is overwhelmed by the electromagnetic Faraday rotation (see, for instance, \cite{demorest04},
and references therein).

However, it is interesting to point out that since the electromagnetic Faraday rotation (\ref{eq:omegaem1}) is
proportional to the square of the wavelength of the radiation (inversely proportional to $\omega^2$), the
situation could be different in the case of short wavelength emission, such as X-rays.

To give a quantitative evaluation, we recall that the polarization position angle (PPA)  can be written in the
form (see \cite{kramer})
\begin{equation}
\alpha_{PPA}(\lambda)=\frac{\alpha_{Faraday}}{2} \equiv \lambda^2 \times RM \label{eq:defPPA1},
\end{equation}
where $\lambda$ is the wavelength, and the rotation measure $RM$ is defined by
\begin{equation}
RM = \frac{e^{3}}{2 \pi m^{2}c^{4}}\int_{sou}^{obs}nB_{%
\parallel }dl.  \label{eq:defRM1}
\end{equation}
The rotation measures of the two pulsars A and B in the double system PSR J0737-3039 have been recently measured
by \cite{demorest04}, and the corresponding values are $-112.3\pm 1.5 \ {rad \ m^{-2}}$ and $-118\pm 12 \ {rad \
m^{-2}}$, respectively. On using these values, we may estimate the PPA of the X-ray emission from PSR J0737-3039
A. On the ground of the recent observations of the X-ray emissions from the double pulsar performed by
\cite{mclau04} and \cite {campana04}, we consider emission at $1 \ keV (\lambda=1.24 \times 10^{-9} m)$, which
gives
\begin{equation}
\alpha_{PPA} (\lambda=1.24 \times 10^{-9} m) \simeq 1.72 \times 10^{-16} {rad}. \label{eq:stimaPPA1}
\end{equation}

In other words, if the other physical conditions are the same, the electromagnetic Faraday effect for X-rays is
some 17 orders of magnitude smaller than for radio wavelengths:   in these conditions, the purely geometrical
frequency independent rotation, i.e. the gravitational Faraday rotation, can become relevant.

%----------------------------Section-------------------------------

\section{Conclusions}

\label{sec:conc}
%----------------------------Section-------------------------------

We have studied the gravitational Faraday rotation on light rays emitted by a pulsar, orbiting another compact
object, such as a pulsar or a black-hole, whose gravitational field, to a reasonable order of approximation, can
be described by the Kerr metric, of which we have considered the suitable weak-field limit.

In this phenomenon, gravito-magnetism plays a leading role, unlike what
happens, for instance, when studying light bending or gravitational time
delay, where it appears as a correction to the gravito-electric contribution.

The effect is a purely geometric one, contrary to the electromagnetic
Faraday rotation, which depends on the frequency of the light rays;
furthermore, it depends on the component of the angular momentum of the
source of the gravitational field along the line of sight. So, when the
latter is not null, a net rotation of the polarization plane is expected.

We have obtained a formula which relates the rotation angle to the orbital
phase of the emitting pulsar, as well as to the parameters describing its
orbit and to the orientation of the angular momentum of the binary companion.

The numerical estimates of the magnitude of this effect, in the context of the double-pulsar system PSR
J0737-3039, show that at radio wavelengths it is overwhelmingly dominated by the electromagnetic Faraday
rotation. Nevertheless, as a result of it's achromatic nature, gravitational Faraday rotation will dominate
electromagnetic Faraday rotation at X-ray energies.  However, in practice, due too its exceedingly small nature,
it is unlikely to be observed in pulsar systems.

Anyway, this effect could be of some relevance for  pulsar-black hole binaries, or accreting black holes
(\cite{connors}).  In such  systems, peculiar modulation of polarization orientation during the orbital motion
may be evidence that an unseen companion is a black hole.

%-------------------------

\section*{Acknowledgments}

%-------------------------
M.L.R acknowledges financial support from the Italian Ministry of University
and Research (MIUR) under the national program `Cofin 2005' - \textit{La
pulsar doppia e oltre: verso una nuova era della ricerca sulle pulsar}.

%--------------------------------------------------------------------------
% The Bibliography
%--------------------------------------------------------------------------

\label{lastpage}

\end{document}